\documentclass[
 reprint,
preprintnumbers,
 amsmath,amssymb,
 aps,
]{revtex4-1}

\usepackage[utf8]{inputenc}
\usepackage{xcolor}
\usepackage{graphicx}

\begin{document}

\title{Measurements of Floquet code plaquette stabilizers}
\author{James R. Wootton}
\affiliation{IBM Quantum, IBM Research Europe -- Zurich}
\date{\today}

\begin{abstract}

The recently introduced Floquet codes have already inspired several follow up works in terms of theory and simulation. Here we report the first preliminary results on their experimental implementation, using IBM Quantum hardware. Specifically, we implement the stabilizer measurements of the original Floquet code based on the honeycomb lattice model, as well as the more recently introduced Floquet Color code. The stabilizers of these are measured on a variety of systems, with the rate of syndrome changes used as a proxy for the noise of the device.

\end{abstract}

\maketitle

\section{Introduction}

Floquet codes are a concept that has recently emerged in quantum error correction~\cite{hastings:21}. The first explicit example was the honeycomb Floquet code~\cite{hastings:21,gidney:21,haah:21,gidney:21b,paetznick:21}, based on Kitaev's honeycomb lattice model~\cite{honey}. The concept was then extended with two additional codes. First was the $e \leftrightarrow m$ automorphism code~\cite{aasen:22}, based on a generalization of honeycomb Floquet code. Next was the Floquet Color code~\cite{brown:22,davydova:22,kesselring:22}, which can be derived from Color codes~\cite{bombin:06}.

\begin{figure}[htbp]
\begin{center}
\includegraphics[width=0.6\columnwidth]{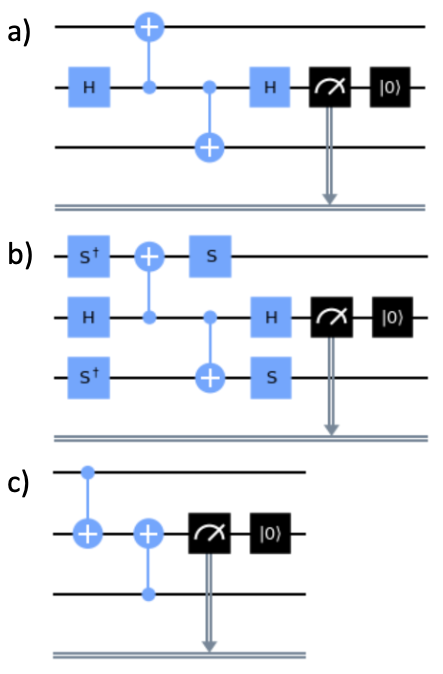}
\caption{Circuits to measure operators for (a) x-links, (b) y-links and (c) z-links. The two vertex qubits for which the parity is measured are at the top and bottom of each circuit. The auxiliary is in the middle. This should be initialized in the $|0\rangle$ state. If it remains in the output state of a previous measurement, the end result will be the XOR of the two.}
\label{circuits}
\end{center}
\end{figure}

Both the original honeycomb Floquet code and the Floquet Color code share similarities in implementation. Specifically, both are based on the measurement of two-body observables corresponding to edges in a hexagonal lattice. This makes them both particularly well suited to the heavy-hexagon architecture of IBM Quantum processors~\cite{chamberland:20}, in which qubits reside on both the vertices and links of a hexagonal lattice. These can then be used as the required code qubits and auxiliary qubits respectively, where the latter are used within the circuits to measure two-body observables (see Fig.~\ref{circuits}).

Most of the available hardware is not sufficient to perform a sensible analysis of the logical subspace. The 27 qubit \textit{Falcon} devices, for example allow only two plaquettes to be realized. Nevertheless, the hardware is sufficient to measure the stabilizers, and assess how they detect noise on the devices. Here we will perform this for the honeycomb Floquet code and Floquet Color codes.

%Additionally we consider the manipulation of twist defects in matching codes~\cite{wootton:majorana}. Though this is not explicitly related to the Floquet codes described above, it is another process that can be performed in hexagonal codes by the measurement of two-body observables.

\section{Honeycomb Floquet codes}

\begin{figure}[htbp]
\begin{center}
\includegraphics[width=0.9\columnwidth]{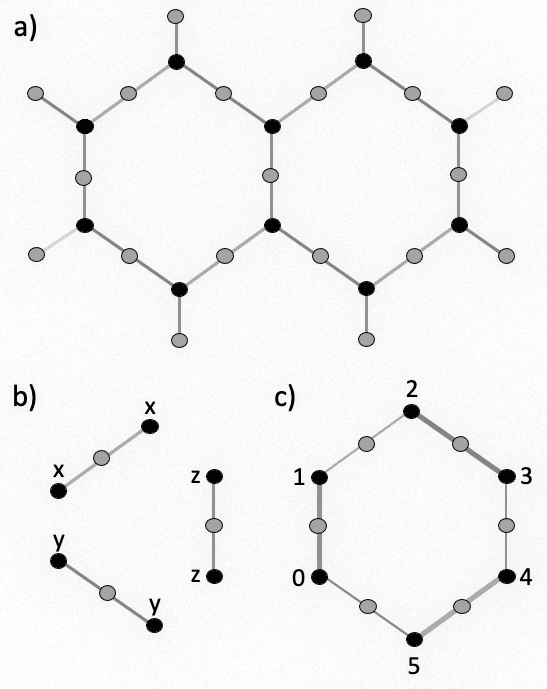}
\caption{(a) Portion of the lattice on which the honeycomb lattice model and derived codes are defined. Vertex qubits are shown as black dots. Grey dots are auxiliary qubits that can be used in the derived codes. (b) Definition of the link operators. (c) Definition of the plaquette operators.} 
\label{code}
\end{center}
\end{figure}

Kitaev's honeycomb lattice is a well-studied model in both condensed matter and quantum information~\cite{honey}. It is defined on a hexagonal lattice with qubits on the vertices, as shown in Fig. \ref{code}. The links are labelled $x$, $y$ and $z$ depending on orientation. Each is assigned a two-body link operator $\sigma^\alpha \otimes \sigma^\alpha$ on the two adjacent vertex qubits, where $\alpha \in \{x,y,z\}$ is the link type.

Important to any study of this model are the plaquette operators, since they commute with all the link operators. Each is defined as the product of link operators around a corresponding plaquette. They can be expressed as
\begin{equation}
W = \sigma_0^x \sigma_1^y \sigma_2^z \sigma_3^x \sigma_4^y \sigma_5^z
\end{equation}
using the numbering of Fig.~\ref{code}(c).

The most obvious means to convert this model to a quantum error correcting code is arguably to perform the two-body parity measurements corresponding to the link operators, treating them as the gauge operators of a subsystem code. The plaquettes then form the corresponding stabilizer generator. Unfortunately, such a code has a trivial logical subspace~\cite{suchara:11}. However, two separate means to carve out a non-trivial subspace were introduced in the last year. One of these was the honeycomb Floquet code, and the other was an approach based on matching codes~\cite{wootton:21}. In both, a particular order is used for the measurement of the gauge operators in order to effectively create a logical subspace.

\begin{figure}[htbp]
\begin{center}
\includegraphics[width=0.9\columnwidth]{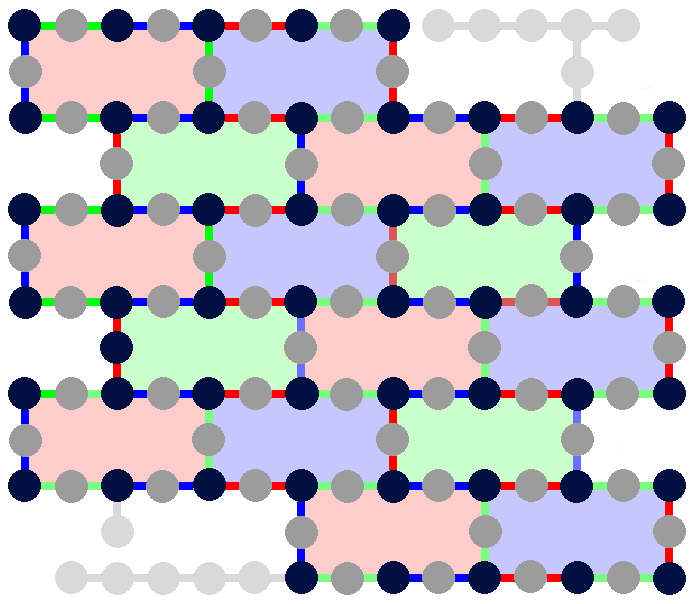}
\caption{Colouring of plaquettes and links for the \texttt{ibm\_washington} device, which is a 127 qubit \textit{Eagle} processor. Qubits shown in black are those of the code, those in grey are auxiliary qubits used during gauge operator measurements, and those in light grey are unused. The plaquettes of the 27 qubit \textit{Falcon} devices and the 65 qubit \textit{Hummingbird} processor are subsets of those shown here. The \textit{Falcon} devices support two plaquettes, corresponding to a red and blue pair. The \textit{Hummingbird} processor supports 8 plaquettes, corresponding to four rows of pairs alternating between red and blue, and green and red.}
\label{hardware}
\end{center}
\end{figure}

For the honeycomb Floquet codes, the measurement schedule is defined using a tri-colouring of the plaquettes and links of the hexgaonal lattice. This is done such that any link of a given colour connects two plaquettes of that colour, and forms part of the boundary of two plaquettes of different colours. The colouring used for the plaquettes of the devices used are shown in Fig. \ref{hardware}

\subsection{Measuring stabilizer changes}

Measurement of the gauge operators is done by measuring the link operators of each colour in succession. We will consider the order red-green-blue by default. The results of any two successive rounds can be used to infer the results for a subset of stabilizers. This is done by simply adding their results modulo 2. For the initial red and green rounds, it is the blue plaquettes for which results can be inferred. This is because the boundary of blue plaquettes is formed by red and green links. After the subsequent round of blue link measurements, results from the green and blue rounds can similarly be used to infer results for the red plaquettes, and so on.

Assuming that the state begins in a product state, the initial measurement of each plaquette will typically give a random result. In the noiseless case, the next measurement would repeat the same result. A mismatch between subsequent measurements of the same plaquette is therefore a detection of an error. To perform such syndrome comparisons on all plaquettes, at least seven rounds of link operator measurements must be performed. It is this minimal seven round case that we will consider here.

We will also consider the case in which resets are not applied after measurement of the auxiliary qubits. The results for each link qubit measurement is then actually the parity of all measurements of that link qubit so far. The first syndrome change for red and blue plaquettes can then be calculated directly from the results of the second instance of the pertinent link operators. For the green plaquettes, only results from the second instance of the blue link measurements are required, but results from the first and third instance of the red link measurements must be combined.

\subsection{Results}

\begin{figure}[htbp]
\begin{center}
\includegraphics[width=\columnwidth]{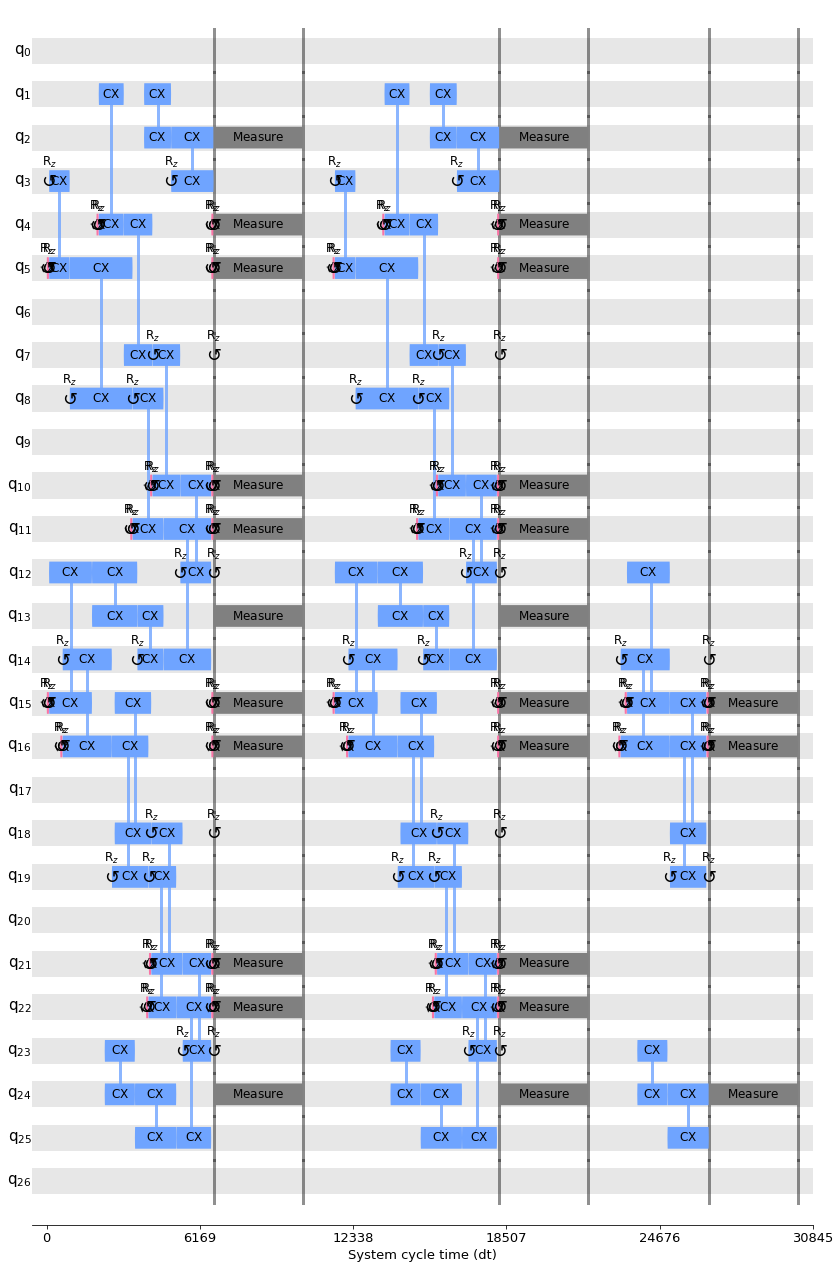}
\caption{A scheduled circuit for \texttt{ibmq\_cairo}. The sample time for this device is $dt=0.22\rm{ns}$. Though circuit does not include dynamical decoupling for clarity, the circuits as run include a pair of \texttt{x} gates during each delay.}
\label{circuit}
\end{center}
\end{figure}

The seven round process is run for both IBM Quantum devices and simulated noise. For the former, an example of a scheduled circuit is shown in Fig.~\ref{circuit} for \texttt{ibmq\_cairo}. The circuits typically settle into three distinct layers. The first consists of the two qubit gates required for the first three rounds of link operator measurement, followed by the measurements on auxilliary qubits for all. This is then repeated for the next three rounds. The final layer is similar, but consists only of the final round of link operator measurements. Note that there is significant idle time for the qubits, caused by both measurement and the scheduling of two qubit gates. The latter is exacerbated by the mismatch in two qubit gate times. In the first layer, this can mitigated to a degree by delaying the first gate for as long as possible for some qubits. This is because idling minimally effects qubits in their initialized state.

For the simulations, only the 27 qubit case is considered. The noise model is a variation of the standard one used for surface code threshold calculations~\cite{stephens:14}, in which everything that can go wrong does so with some a probability $p$. Specifically, bit flips after reset and prior to measurement occur with probability $p$, and depolarizing noise is applied to each two qubit gate and during idle times with probability $p$. The idling error is applied during measurement to the unmeasured qubits. It is also applied to each qubit at the beginning of each layer, as a proxy for the idling during two qubit gates.

Whether for quantum devices or simulation, the results are processed to calculate a syndrome change detection rate for each plaquette. This is the fraction of shots for which a change in value was detected for the two measurements of the corresponding plaquette. Since such a change in value occurs only in the presence of errors, this rate will be zero for $p=0$. It can then be expected to converge to the maximally random value of $1/2$ as $p$ increases. The value found for quantum devices will give some sense of how noisy the syndrome extraction process is.

For comparison of IBM Quantum devices with simulations, we will calculate an average of the noise probabilities from the devices obtained in benchmarking. Specifically we consider the following probabilities.
\begin{itemize}
\item Preparation noise: The probabilities \texttt{prob\_meas1\_prep0} for each qubit.
\item Measurement noise: The measurement error probability for each qubit.
\item CX noise: The CX error probabilities.
\item Idle noise: For each qubit, the error probability for an identity gate over the timescale of measurement and the longest CX gate. This is extrapolated from the probability $p_{\rm{id}}$ of an error over the standard timescale $t_{\rm{id}}$ of an identity gate~\cite{wootton:21}.
\end{itemize}
The mean and standard deviation of this set of probabilities is shown for the devices in table below, along with the the quantum volume~\cite{cross:20}. These values are derived from the most recent benchmarking data at the time when the results were taken.

\begin{center}
\begin{tabular}{|c c c c|} 
 \hline
 Device & $\langle p \rangle$ & $\sigma$ & QV \\ [0.5ex] 
 \hline
 \texttt{ibm\_hanoi} \,&\, $1.32\,\%$ \,&\, $1.39\,\%$ \,&\, 64 \\ 
 \hline
 \texttt{ibm\_auckland} \,&\, $1.50\,\%$ \,&\, $1.71\,\%$ \,&\, 64 \\
 \hline
 \texttt{ibmq\_kolkata} \,&\, $1.76\,\%$ \,&\, $3.14\,\%$ \,&\, 128 \\
 \hline
 \texttt{ibm\_peekskill} \,&\, $1.81\,\%$ \,&\, $2.37\,\%$ \,&\, - \\
 \hline
 \texttt{ibmq\_montreal} \,&\, $2.26\,\%$ \,&\, $2.09\,\%$ \,&\, 128 \\
 \hline
 \texttt{ibmq\_mumbai} \,&\, $3.00\,\%$ \,&\, $1.73\,\%$ \,&\, 128 \\
 \hline
 \texttt{ibmq\_brooklyn} \,&\, $3.02\,\%$ \,&\, $3.33\,\%$ \,&\, 32 \\
 \hline
 \texttt{ibmq\_washington} \,&\, $3.13\,\%$ \,&\, $5.39\,\%$ \,&\, 64 \\
  \hline
 \texttt{ibmq\_toronto} \,&\, $4.72\,\%$ \,&\, $6.37\,\%$ \,&\, 32 \\
 [1ex] 
 \hline
\end{tabular}
\end{center}

\begin{figure}[htbp]
\begin{center}
\includegraphics[width=\columnwidth]{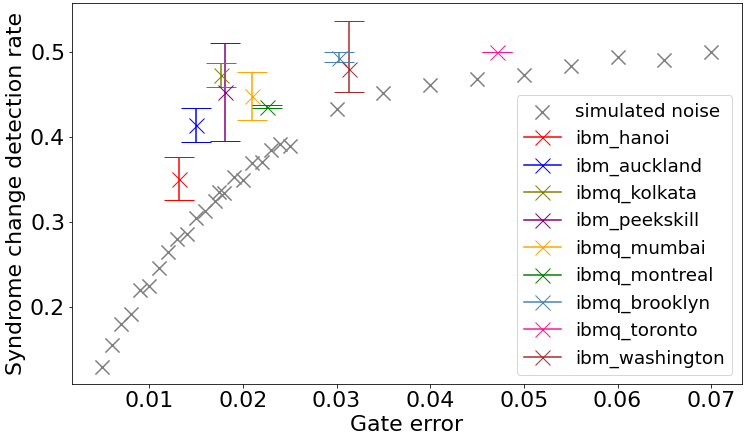}
\caption{The values of the syndrome change detection rate plotted against noise strength ($p$ for simulated runs and $\langle p \rangle$ for quantum hardware). For simulated results, the mean syndrome change detection rate over all plaquettes is shown. For results from quantum hardware, the mean, maximum and minimum over all plaquettes are shown. Specifically, the central point is the mean, and the bars extend down to the minimum and up to the maximum.}
\label{results}
\end{center}
\end{figure}

The results are shown in Fig.~\ref{results}. Results from both simulations and quantum hardware show the expected behaviour as the error rate increases. For \texttt{ibmq\_toronto}, the relatively high noise rate results in completely random syndrome change detection rates. For all other cases, however, results depart significantly from the maximally random value, with ever lower average error rates achieving ever lower syndrome change detection rates. The best results are comparable with a gate error of $p=2\%$ for the simple noise model used in the simulations.

\section{Floquet Color codes}

\subsection{Measuring stabilizer changes}

The Floquet Color codes are also defined using the tri-colouring of the plaquettes and links of the hexagonal lattice shown in Fig. \ref{hardware} For each link we assign two parity operators: $\sigma^\alpha \otimes \sigma^\alpha$ for $\alpha \in \{x,z\}$. For each plaquette we define two plaquette operators,
\begin{equation}
W^\alpha = \sigma_0^\alpha \sigma_1^\alpha \sigma_2^\alpha \sigma_3^\alpha \sigma_4^\alpha \sigma_5^\alpha, \,\,\,\, \alpha \in \{x,z\},
\end{equation}
using the numbering of Fig.~\ref{code}(c).

Note that the set of parity operators here do not commute with the plaquette operators. This code therefore does not correspond to a subsystem code in which the link operators are gauge operators and the plaquette operators are stabilizers. Nevertheless, the roles they play are equivalent.

Each round of measurements consists of measuring a single link type of a single colour of links, alternating between the two link types and through the three colours from round to round. A complete cycle therefore takes 6 rounds~\cite{brown:22,davydova:22,kesselring:22}.

The measurement outcome for a red $W^\alpha$ plaquette operator can be deduced from the $\alpha$-link measurements for both green and blue links, and therefore twice per cycle. However, note that the red $W^\alpha$ plaquette operator does not commute with the $\neg \alpha$-link operators for red links. The outcome of the red $W^\alpha$ plaquettes following the measurement of red $\neg \alpha$-link operators will therefore produce random results, uncorrelated with any previous results. However, the subsequent red $W^\alpha$ plaquettes will not be so disturbed, and therefore should repeat the same result. A mismatch between subsequent measurements of the same $W^\alpha$ on the same plaquette is therefore a detection of an error. These syndrome comparisons form the basis of error detection in the code, since any measured syndrome change is a signature of an error. Equivalent arguments can be made for the blue and green plaquettes.

As with the honeycomb Floquet code, the scheduled circuit will settle into distinct layers in which each consist of three rounds of link operator measurements. This is because such sets of three rounds directly measure different auxiliary qubits, allowing the measurement gates to be performed relatively simultaneously. Each set of three rounds performs both types of plaquette measurement on both types of plaquette. A single six-round cycle is therefore sufficient to determine syndrome changes for several types of $W^\alpha$. However, due to the scheduling of the non-commuting link measurements, ten rounds are required for syndrome changes of all~\cite{davydova:22,kesselring:22},. Note that we do not consider changes in syndrome from the initialized value, which is possible for some of the $W^z$ stabilizers. This is because we are interested in the typical syndrome changes within a computation, rather than this edge case.

\subsection{Results}

The ten round process is run for both IBM Quantum devices and simulated noise. The layers within the scheduled circuits are visually similar to those of the honeycomb Floquet code shown in Fig.~\ref{circuit}, and so have similar considerations concerning idling times. The same noise model as before is therefore used for the simulations. The quantities measured in these runs are the syndrome change detection rate for each plaquette $W^{\alpha}$.

The comparison noise probabilities are calculated in the same way as before. However, since results were taken on different days, the benchmarking data from which the values are derived are not the same. The relevant set of probabilities for these results is therefore shown for the devices in table below.

\begin{center}
\begin{tabular}{|c c c c|} 
 \hline
 Device & $\langle p \rangle$ & $\sigma$ & QV \\ [0.5ex] 
 \hline
 \texttt{ibm\_hanoi} \,&\, $1.40\,\%$ \,&\, $1.63\,\%$ \,&\, 64 \\ 
 \hline
 \texttt{ibmq\_kolkata} \,&\, $1.47\,\%$ \,&\, $1.80\,\%$ \,&\, 128 \\
 \hline
 \texttt{ibm\_auckland} \,&\, $1.82\,\%$ \,&\, $3.36\,\%$ \,&\, 64 \\
 \hline
 \texttt{ibm\_peekskill} \,&\, $2.16\,\%$ \,&\, $3.06\,\%$ \,&\, - \\
 \hline
 \texttt{ibmq\_montreal} \,&\, $2.26\,\%$ \,&\, $2.09\,\%$ \,&\, 128 \\
 \hline
 \texttt{ibmq\_mumbai} \,&\, $2.38\,\%$ \,&\, $2.04\,\%$ \,&\, 128 \\
 \hline
 \texttt{ibmq\_brooklyn} \,&\, $2.78\,\%$ \,&\, $2.96\,\%$ \,&\, 32 \\
  \hline
 \texttt{ibm\_washington} \,&\, $3.10\,\%$ \,&\, $4.78\,\%$ \,&\, 64 \\
  \hline
 \texttt{ibmq\_montreal} \,&\, $4.36\,\%$ \,&\, $7.82\,\%$ \,&\, 128 \\
 [1ex] 
 \hline
\end{tabular}
\end{center}

\begin{figure}[htbp]
\begin{center}
\includegraphics[width=\columnwidth]{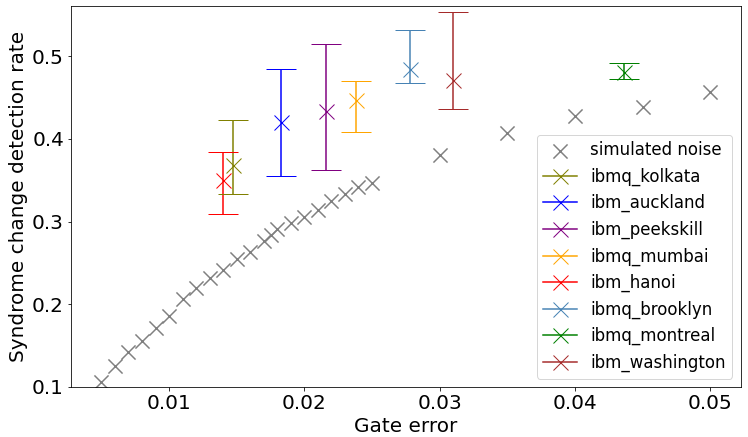}
\caption{The values of the syndrome change detection rate plotted against noise strength ($p$ for simulated runs and $\langle p \rangle$ for quantum hardware). For simulated results, the mean syndrome change detection rate over all plaquettes is shown. For results from quantum hardware, the mean, maximum and minimum over all plaquettes are shown. Specifically, the central point is the mean, and the bars extend down to the minimum and up to the maximum.}
\label{results_bk}
\end{center}
\end{figure}

The results are shown in Fig.~\ref{results_bk}. Results from both simulations and quantum hardware show the expected behaviour as the error rate increases. For the simulations, we see that the syndrome detection rate for a given $p$ is lower than that for the honeycomb Floquet codes. Nevertheless, results from quantum hardware show similarities between the two codes For example, devices with $\langle p \rangle \approx 2 \%$ have a syndrome detection rate of just over $0.4$ for both.

%\section{}

\section{Conclusions}

Here we report results from preliminary implementations of two forms of Floquet code. The results can be used to get a sense of both how well current hardware can implement such codes, as well as how well the codes perform.

From the perspective of the hardware, we find that current IBM Quantum devices are already up to the task of performing syndrome measurements for these codes. Specifically, the syndrome change detection rate is significantly below the maximally random value of $0.5$ in almost all cases. Nevertheless, it is clear that the noise level has only just reached the point where a clear signal is possible. Indeed, the clarity of the signal was marginal for the similar study performed in \cite{wootton:21}, which was for a different code but with much of the same hardware. Much of the improved performance seen in the current work is dependent on suitable use of dynamical decoupling. It will therefore not only be interesting to see progress made in future devices, but also to see how much further current devices can be pushed.

From the perspective of the codes, the results do not clearly distinguish between the two. For the simulations, the syndrome change detection rates for the Floquet Color code are slightly lower than for the honeycomb Floquet code. However, this advantage does not also appear for runs on quantum hardware, at least to an appreciable degree.

This work is not intended to give a definitive assessment of either the codes or the hardware. Instead, these results are a reference by which we can compare future progress.

The source code and data for all results in this paper is available at~\cite{data4papers}.
\\

\emph{Acknowledgements:} This work was partially supported by IARPA under LogiQ (contract W911NF-16-1-0114). All statements of fact, opinion or conclusions contained herein are those of the authors and should not be construed as representing the official views or policies of the US Government. The software used was created as part of a project supported as a part of NCCR SPIN, a National Centre of Competence in Research, funded by the Swiss National Science Foundation (grant number 51NF40-180604). The author would like to thank Benjamin Brown and Markus Kesselring for discussions of their work.
%IBM, the IBM logo, and ibm.com are trademarks of International Business Machines Corp., registered in many jurisdictions worldwide. Other product and service names might be trademarks of IBM or other companies. The current list of IBM trademarks is available at \url{https://www.ibm.com/legal/copytrade}.

\bibliography{references}

\end{document}